# Three-Dimensional Imaging of Individual Point Defects Using Selective Detection Angles in Annular Dark Field Scanning Transmission Electron Microscopy

Jared M. Johnson, Soohyun Im, and Jinwoo Hwang[*]

Department of Materials Science and Engineering, The Ohio State University, Columbus, OH 43212, USA

[*]To whom the correspondence should be addressed. Electronic mail: hwang.458@osu.edu

**Abstract:**

We propose a new scanning transmission electron microscopy (STEM) technique that can realize the three-dimensional (3D) characterization of vacancies, lighter and heavier dopants with high precision. Using multislice STEM imaging and diffraction simulations of $\beta$-$Ga_2O_3$ and $SrTiO_3$, we show that selecting a small range of low scattering angles can make the contrast of the defect-containing atomic columns substantially more depth-dependent. The origin of the depth-dependence is the de-channeling of electrons due to the existence of a point defect in the atomic column, which creates extra "ripples" at low scattering angles. We show that, by capturing the de-channeling signal with narrowly selected annular dark field angles (*e.g.* 20-40 mrad), the contrast of a column containing a point defect in the image can be significantly enhanced. The effect of sample thickness, crystal orientation, probe convergence angle, and experimental uncertainty will also be discussed. Our new technique can therefore create new opportunities for highly precise 3D structural characterization of individual point defects in functional materials.

## 1. Introduction

Recent advances in transmission electron microscopy (TEM), such as aberration correction, have enabled imaging of materials with sub-Angstrom resolution, and delivered significant impact to many disciplines of science. The next phase of electron microscopy must involve fully understanding the signal-generating mechanisms and developing new ways to extract maximal information about the materials' structure from the data. One example is the characterization of point defects in materials. Point defects have played a major role in tuning the electronic and optical properties of semiconductors over the last few decades, and the ability to control individual point defects will continue to be an essential part of the development of the next-generation functional materials (*e.g.* [1, 2]). To control individual point defects, it is required to determine the exact location, distribution, segregation, or clustering of the point defects, and understand how they affect the local structure and functional properties. While there are indirect spectroscopic methods that can estimate point defect concentrations (*e.g.* [3]), establishing the direct relationship between the detailed structural aspects of point defects and the properties of the material has been challenging. With unmatched spatial resolution, TEM-based techniques, including scanning TEM (STEM) [4-10] and electron energy loss spectroscopy [11-14], have been able to detect the impurity atoms in crystals. The remaining important task is to obtain the exact depth information of the point defect along the beam direction in TEM. The depth information is required to determine the 3D positions of individual point defects, which, for example, is necessary to reveal their clustering or segregation at surfaces or interfaces that may decide whether they are electronically active or not. The aforementioned 3D information is also important to determine how the point defects are related to other important aspects in the structure, such as bonding distances, local lattice distortion, and extended defects in crystals. Acquiring the true 3D information of atomic scale defects using electron microscopy can therefore significantly advance our knowledge in many areas of materials research.

Efforts have been made to acquire the depth information of individual atoms in the past using STEM [6, 15-17]. In STEM, increasing the probe convergence angle should ideally increase the depth resolution and therefore STEM imaging using highly converged probes should be able to provide 3D atomic scale information. However, even with the largest convergence half angle currently possible (~30 mrad), the depth resolution still remains above ~ 5 nm, far greater then the size of an atom. To increase the limit of depth information in STEM, focal series imaging with confocal configurations have been studied [18-22]. An alternative approach is to use the electron channeling effect in the crystal. When a converged probe enters a crystal matrix, the probe intensity oscillates as the electrons pass through the sample along the depth direction [6, 15, 23, 24]. The oscillation results in the difference in the amount of electron scattering that occurs at each atom position along the depth direction. Therefore the exact understanding of how probe channeling occurs in a certain crystal should allow for identifying the depth position of the atomic scale object based on the final image intensity. In other words, the probe oscillation can provide the ability to "resolve" the depth information of the atomic scale object. This method has been used to visualize the atomic structure buried at interfaces [25], and to determine the depth positions of the individual heavier dopants in a crystal [6, 26]. One of the key advantages of this method is that it only needs a single scan of the probe on the sample, hence it can prevent the complications that arise due to multiple scans of the probe in focal series imaging, such as sample drift or radiation damage. Furthermore, while a thin TEM sample is still necessary, the sample can be as thick as a few to ten nanometers, depending on the material [5, 6, 26, 27]. Recent reports have demonstrated that preparing such high quality thin TEM samples from a bulk (or thin film) would require the mechanical wedge polishing technique [6, 26, 27], rather than the use of the more popular focused ion beam method.

In general, the 3D characterization of point defects using electron channeling information should satisfy two main experimental requirements: (*i*) a fully quantitative comparison between the experimental and theoretical image intensities, and (*ii*) the exact knowledge about the thickness of the sample area that is being scanned. The former has been realized using the quantitative STEM technique [28] based on the exact calibration of the annular dark field (ADF) detector and multislice STEM simulations, and the latter is possible by simultaneously acquiring position averaged convergent electron beam diffraction patterns [29] during imaging. Recently, using probe channeling information combined with quantitative STEM, Hwang et al. have shown that the depth of individual Gd dopant atoms in the Sr columns of $SrTiO_3$ was determined with depth uncertainty less than ±1 unit cell [6]. Among the atomic ($Z$) number based heavier dopant imaging, this has shown the smallest $Z$-number ratio between the dopant ($_{64}$Gd) and the host ($_{38}$Sr) so far. The authors also found that ultimately the dopant depth precision is limited by the experimental errors, which may include sample drift, surface roughness, scan distortion, and Poisson noise. A following report by Zhang et al. [26] has shown that the depth precision can be further increased using the signals from multiple ADF detection angles, as the dopants in different depths may scatter in different angles. These results have a couple of important implications: first, the ADF detection angles can be optimized to maximize the depth dependence of the point defect contrast in the image, and second, theoretical calculation of the image intensity and probe channeling behavior should play a critical role in the prediction and interpretation of the point defect imaging.

While detecting heavier dopants has been possible using $Z$-contrast imaging combined with probe channeling information, detecting vacancies and lighter dopants, including their 3D positions, has remained difficult, as the exact understanding of the contrast of the atomic columns including them has been lacking. In this paper, we report a new STEM imaging mode that can realize the 3D characterization of vacancies, lighter and heavier dopants. Using multislice STEM image and diffraction simulations, we show that selecting a small range of detection angles (with about 10 mrad in width) can make the contrast of the defect-containing atomic columns substantially more depth-dependent. We also show that the low angle ADF (LAADF) signals (*e.g.* 20-40 mrad) are particularly sensitive to the depth of vacancies and lighter dopants, while high angle ADF (HAADF) (*e.g.* 100-150 mrad) signals are more sensitive to the depth of heavier dopants. We will also discuss potential ways to realize such selective detection angles in real experimental setups, and consider the maximum experimental error range that can enable indisputable identification of the depth positions of individual point defects.

We used $β$-Ga$_2$O$_3$ and SrTiO$_3$ for this simulation study. $β$-Ga$_2$O$_3$ is a transparent conductive oxide (TCO) with an ultra wide band gap (UWBG) of ~4.9 eV [30]. $β$-Ga$_2$O$_3$ is effectively a direct band gap material [31] that shows transparency up to ultraviolet (UV) regions [32] and has a high break down voltage [33]. It is also available as a high quality single crystal substrate that can be prepared using conventional bulk growth methods [34]. Due to these unique advantages, *β*-Ga$_2$O$_3$ has recently gained significant attention for high-performance, high-efficiency UWBG applications, such as UV transparent electrodes, high-power and high-voltage field effect transistors, and photodetectors (*e.g.* [35, 36]). As in other TCOs, the transport and doping properties of *β*-Ga$_2$O$_3$ is significantly affected by the types of point defects in it [37]. Undoped *β*-Ga$_2$O$_3$ typically shows *n*-type behavior. Unlike in some other TCOs, however, the oxygen vacancies may not be the direct reason for the intrinsic *n*-type behavior because they are deep donors in *β*-Ga$_2$O$_3$ [31, 38, 39]. Instead, unintentionally doped (UID) impurity atoms, such as Si [40] and H [38, 41], may be the source of the *n*-type behavior. Such UID impurities can also have close relationship to cation (Ga) vacancies. Understanding the details of the point defect complexes and how they are connected to the local lattice distortion [42] and electronic properties can therefore open new possibilities of tuning the material's properties for many important UWBG applications. From a structural point of view, the low-symmetry (monoclinic) structure of *β*-Ga$_2$O$_3$ provides a unique opportunity to study the effects of orientation on probe channeling and dopant contrast in the STEM image. Here we studied the change in the LAADF and HAADF intensities of the columns in *β*-Ga$_2$O$_3$ containing cation vacancies and substitutional Si and Nd [43] impurity atoms. Some of the important results that we acquired, such as the ripple effect at low scattering angles, were also confirmed using a more widely studied higher-symmetry crystal, SrTiO$_3$.

## 2. Simulation Details

All STEM ADF image simulations were performed using multislice algorithm [44] with thermal diffuse scattering (TDS) using the frozen phonon approximation at 298 K. The root mean square (RMS) deviation values of the thermal vibration of *β*-Ga$_2$O$_3$ were taken from Ahman et al. [45]. For each image simulation with TDS, ten images, each having twenty frozen phonon configurations, were first simulated, and then averaged to increase statistical reliability, unless stated otherwise. This should be statistically equivalent to simulating one image with two hundred phonon configurations, but ten times faster. The uncertainty of the averaging is presented with the standard deviation of the mean (SDOM) [46]. The image simulations used 1024 by 1024 real and reciprocal space sampling, with the optical parameters of our probe corrected FEI Titan STEM ($Cs_3$=0.002, $Cs_5$=1.0). Two convergence half angles, 9.6 and 20.0 mrad, were tested. The convergent beam electron diffraction (CBED) simulations were performed with and without TDS. For CBEDs with TDS, 1,000 frozen phonon configurations were used.

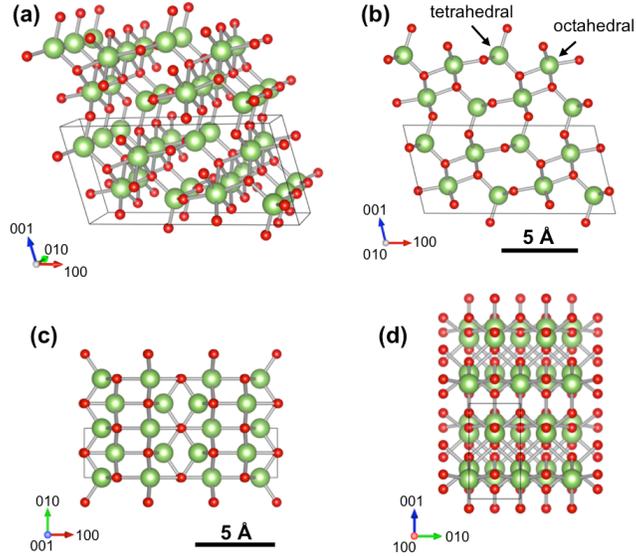

**Figure 1.** Crystal structure of $\beta$-$Ga_2O_3$ in a (a) perspective view, along (b) $[010]_m$, (c) $[001]_m$, and (d) $[100]_m$ view. Green spheres are Ga atoms and red spheres are oxygen atoms.

$\beta$-$Ga_2O_3$ has space group $C\ 2/m$ (monoclinic) (Fig. 1a). Among the three <100> axes, we selected the monoclinic [010] (Fig. 1b) and [001] (Fig. 1c) (therefore $[010]_m$ and $[001]_m$, respectively) as the imaging orientations for this study. $[100]_m$ was not used because the atomic columns in the image would overlap significantly and may complicate the analysis (Fig. 1d). As shown in Fig. 1, the $[010]_m$ orientation will have the largest gap between the Ga atoms in the image (interatomic distance ~3.3 Å), but it will have a relatively short (3.04 Å) interatomic distance along the column (beam direction). The $[001]_m$ orientation has tighter distances between the columns in the image (Fig. 1c), but has longer interatomic distance along the beam direction (5.8 Å). As we will discuss in Section 3.4, the interatomic distance along the beam direction significantly affects the probe channeling oscillation and therefore affects the usable sample thickness range in point defect imaging. Ga atoms exist in two different positions in $\beta$-$Ga_2O_3$: one in the octahedral and the other in the tetrahedral (Fig. 1b). While such distinction in Ga positions may be important for studying point defect complexes (*e.g.* [39]), it carries little importance for the present study as we found that those two positions do not show a significant difference in point defect contrast or probe channeling oscillation. Therefore we only show the results from the point defects positioned at the octahedral Ga positions in this paper. Sample thicknesses of 33 and 55 Å were used for $[010]_m$ orientation, and 60 and 119 Å were used for $[001]_m$ orientation.

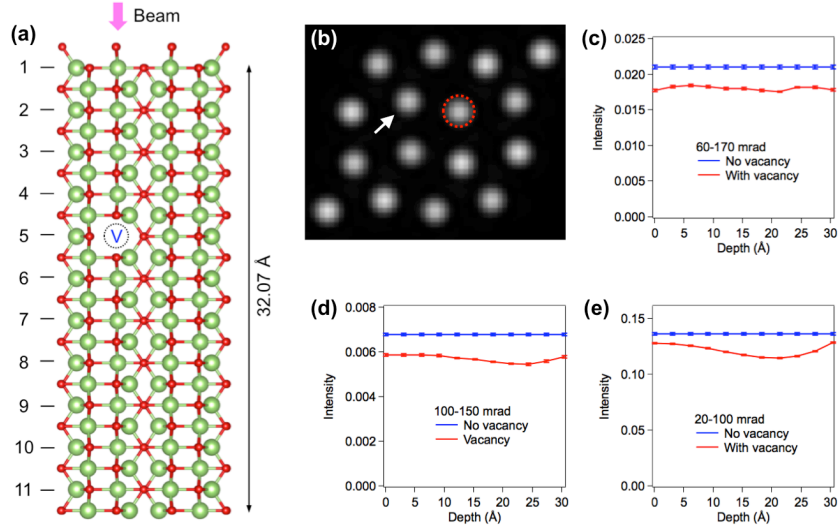

**Figure 2**. (a) Schematic of *β*-$Ga_2O_3$ used in the simulation along the $[010]_m$ beam direction. Vacancies were placed one-by-one at the positions indicated along the depth, 1 to 11. (b) An example simulated image using a HAADF detection angle (60-170 mrad). The arrow indicates the column containing the vacancy. The red dotted circle shows the area that we averaged to get the intensity of the column. The column intensity as a function of the vacancy depth position (red curve) was plotted against the intensity of the column without a vacancy (blue curve) for ADF angles (c) 60-170 mrad, (d) 100-150 mrad, and (e) 20-100 mrad. The error bars indicate the uncertainty (SDOM) generated from the averaging of ten simulated images, each of which used twenty frozen phonon configurations at 298 K. Probe convergence half angle = 9.6 mrad, and sample thickness = 33 Å.

## 3. Results

### 3.1. Achieving depth-dependent vacancy contrast using LAADF STEM

We first show the image simulation along the $[010]_m$ orientation (Fig. 2). Simulations were carried out using different ADF detection angles for single vacancies placed in the Ga atomic column (Fig. 2a) at each depth position (1-11). An example of the resulting HAADF STEM image is shown in Fig. 2b. The white arrow indicates the column containing the vacancy. The dotted red circle indicates the area where we averaged the intensity for the column. This column-averaged intensity is especially useful for the analysis of experimental images, as it tends to reduce the random experimental errors [6, 47]. Figures 2c to 2e show the column-averaged intensity as a function of the depth position of the vacancy along the column (red curve), as compared to the intensity of the column with no vacancy (blue curve). Figure 2c shows intensity vs. depth using the typical HAADF detection angle range, which is also the range used in previous quantitative STEM works [6, 26, 28] (the actual range used in the experiment was 60-380 mrad, but there is almost no intensity beyond 170 mrad). While the vacancy-containing column intensity is significantly lower than that of normal columns (without vacancy), the intensity profile is essentially independent of the vacancy depth position. This is in fact a problem, because it will be impossible to distinguish the vacancy in the bulk from the vacancy on the surface, which is not even distinguishable from the atomic scale surface roughness that may be present on the sample surface. The trend shown in Fig. 2c is also different from the heavier atom (Gd) in a lighter host (Sr) observed previously [6, 26], where the column intensity increases monotonically as a function of the dopant depth position. The dissimilar trends between the heavier dopant and vacancy is due to the difference in the way they affect the probe channeling oscillation, which we will discuss in Section 3.6. On the other hand, when using a smaller angle range in HAADF, 100 to 150 mrad (Fig. 2d), the profile becomes slightly more depth-dependent, showing a small decrease at ~24 Å depth. The amount of the intensity decrease is about 7%, which may be detectable depending on the amplitude of experimental errors. However, when LAADF angles (20-100 mrad) are used, the intensity decrease is more prominent than in HAADF. At ~20 Å depth, the intensity drops to about 10% of the intensity of the cases where the vacancies are positioned at either surface. This implies that, unlike detecting heavier dopants using HAADF imaging [5, 6, 16, 22, 26], vacancies in the bulk can be more easily detected using LAADF angles. In Section 3.2, we will show that the contrast of the defect-containing column can be further enhanced if we choose a smaller detection angle range in LAADF. But first, we will show why LAADF signal is more sensitive to vacancy depth position using CBED simulation.

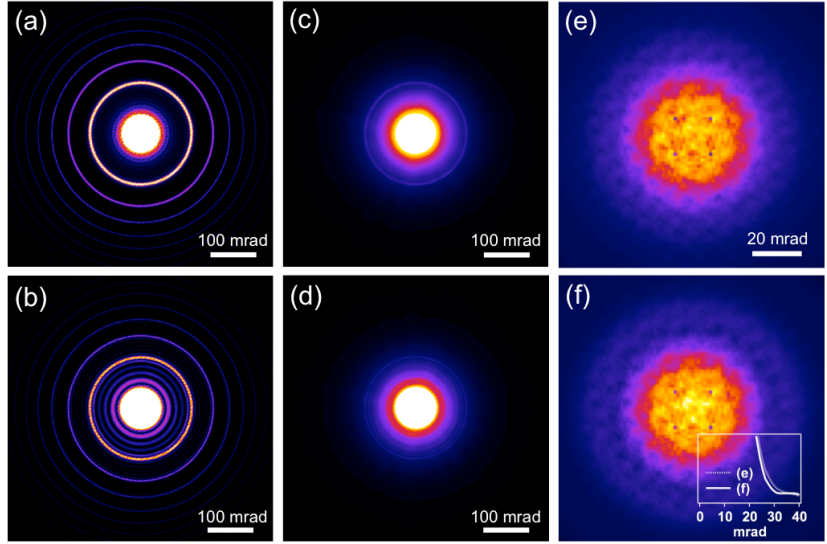

**Figure 3**. CBED patterns simulated with a probe positioned at the octahedral Ga column with (a) no vacancy and (b) a vacancy at the 6$^{th}$ position from the top surface without TDS. (c) and (d) show the same simulations but with TDS for (a) and (b) cases, respectively. (e) and (f) are the zoomed-in versions of (c) and (d), respectively, to show the intensities at low scattering angles. The patterns (a) to (d) were presented in linear scale with the intensity saturated at the center to make the higher angle scattering intensities more visible. The patterns in (e) and (f) were presented with $\gamma = 0.5$. The inset in (f) shows the annular averaged intensities of the 2D patterns in (e) and (f) to show the differences between them in linear scale. The $x$-scale of the inset graph matches the $x$-scale of the 2D pattern in (f). Probe convergence half angle = 9.6 mrad, and sample thickness = 33 Å.

The cause of the LAADF's dependence on the vacancy depth can be found in CBED patterns simulated with the probe positioned at the Ga column (Fig. 3). We first look at the CBED patterns without TDS for columns (a) without a vacancy and (b) with a vacancy positioned at the 6$^{th}$ position from the top (15 Å depth) along [010]$_m$ orientation. The two patterns show similar Higher Order Laue Zone (HOLZ) intensities beyond ~100 mrad, but show remarkably different patterns at lower angles (20-100 mrad). The extra "ripples" at lower angles shown in Fig. 3b must be the result of the de-channeling of the electron due to the existence of the vacancy in the column. It has been well known that any abnormality or defect (*e.g.* dislocation) in a crystal can result in the strong de-channeling of the electron, which can be captured using LAADF [48, 49]. The result here indicates that similar de-channeling can happen by the perturbation from the vacancy. When TDS is used in the simulations for the columns (c) without a vacancy and (d) with a vacancy, we see in both most of the HOLZ patterns are diffused, but the low angle ripple effect caused by the vacancy still remains. Figures 3e and 3f show that the difference between the patterns with and without vacancy caused by the ripple effect is most significant in the angle range between 20 and 40 mrad. The CBED simulation result implies that, if the ADF detection angle is limited to 20-40 mrad, the contrast of the column containing a vacancy can be highly enhanced.

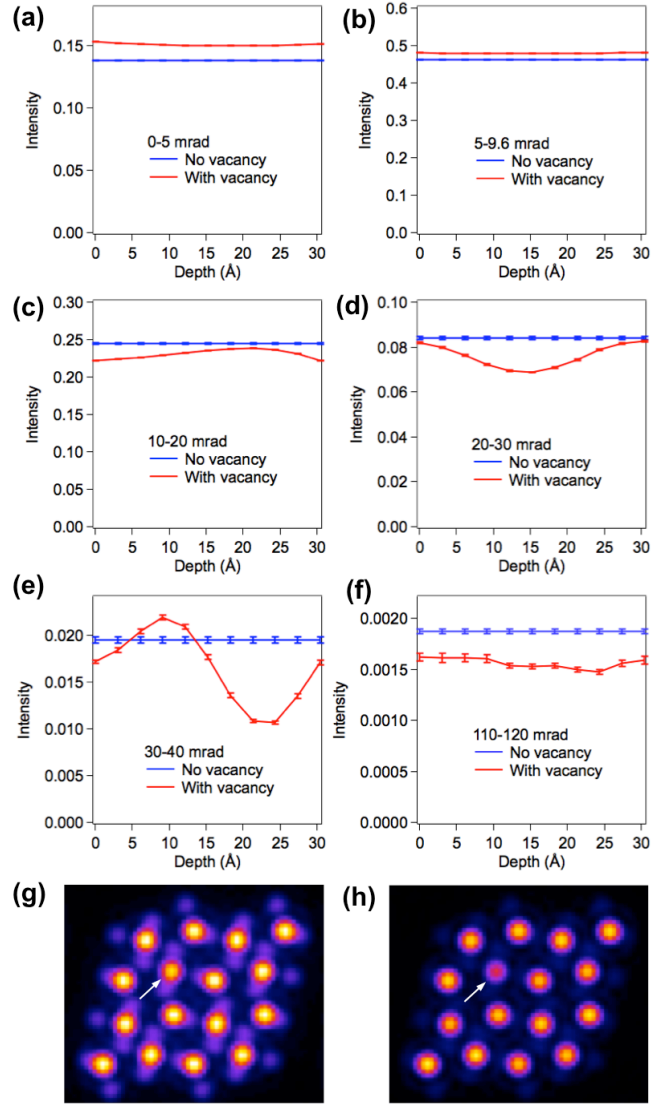

**Figure 4**. Column intensity as a function of vacancy depth position (red curve) compared to the column intensity without vacancy (blue curve) for detection angles (a) 0-5 mrad (bright field), (b) 5-9.6 mrad (annular bright field), (c) 10-20 mrad, (d) 20-30 mrad, (e) 30-40 mrad, and (f) 110-120 mrad, for imaging with $[010]_m$ orientation. (g) is the simulated image with a vacancy placed at the 6$^{th}$ position (15 Å depth) from the top (see Fig. 2a) with 20-30 mrad detection angle. (h) is the simulated image with vacancy placed at the 9$^{th}$ position (25 Å depth) from the top with 30-40 mrad detection angle. All simulations used TDS and probe convergence half angle of 9.6 mrad, and sample thickness = 33 Å.

3.2. 3D imaging of individual vacancies using selective LAADF ranges

Here we show that accurate determination of the 3D position of individual vacancies should be possible using multiple, narrowly selected LAADF angles. Figures 4a to 4f show the intensities of the columns that contain the vacancy as a function of vacancy depth from the top surface (red curves) as compared to the intensity of the column without a vacancy (blue curves), all with 9.6 mrad probe convergence half angle but with different ADF detection angles. The result shows that, the bright field (Fig. 4a) and annular bright field (Fig. 4b) intensities show essentially no dependence on the vacancy depth, meaning the identification of bulk vacancy will be difficult if these angles are used. However, consistent to the CBED results shown in Fig. 3, the most variation in column intensity occurs when the detection angles are 20-30 mrad (Fig. 4d) and 30-40 mrad (Fig. 4e). With the 20-30 mrad angle range, the vacancy in the bulk can drop the column intensity to about 20% of the intensity of the column without a vacancy (blue curve) or the column with surface vacancies (both ends of the red curve), which means the accurate identification of bulk vacancies should be possible. However, the 20-30 mrad angle shows a symmetric profile along the vacancy depth, so distinguishing between the vacancy at the upper part and the vacancy at the lower part of the sample, for example, one placed at the $3^{rd}$ position (6 Å depth) and the other at the $8^{th}$ position (22 Å depth) from the top, may be difficult. However, incorporating the information from the 30-40 mrad angles can resolve this problem, because it shows significantly different intensities for those positions. The 30-40 mrad range in fact shows the most significant decrease in the intensity when the vacancy is at the $9^{th}$ position from the top, where the intensity is almost half the intensity of the column without a vacancy (see Fig. 4h). It is also worth pointing out that, with 30-40 mrad angles, the intensity of the vacancy-containing column can even be higher than that of the column without vacancy, which happens near ~10 Å depth. This "contrast reversal" indicates that the signal intensity in this angle range is dominated by the channeling ripple effect shown in Fig. 3. In addition, the 20-30 mrad range may also be useful to study how the point defect changes the nearby Ga-O bonding, as the diffraction contrast reveals the oxygen positions well (Fig. 4g). The high diffraction contrast of oxygen columns at LAADF must be due to the complexity of the $\beta$-$Ga_2O_3$ structure that generates many extra diffraction peaks. This is not the case, for example, in the image simulated for a higher symmetry crystal such as $SrTiO_3$, where the oxygen columns are nearly invisible when the 20-30 mrad detection angle range is used. The HAADF angle, 110-120 mrad (Fig. 4f), basically shows the same trend as the 100-150 mrad range (Fig. 2d), which indicates minimal dependence on the vacancy depth position. The result shows that the detection of the vacancy in the bulk with precise depth information should be possible by combining the information of multiple selective LAADF signals. We have proposed the idea of combining information from multiple ADF detectors to increase the depth precision in Zhang et al. [26], but the present work reveals that the depth sensitivity of the column intensity can be greatly increased by limiting the width of the annular detector to about 10 mrad and by optimizing the detection angle range depending on the type of point defects. Using LAADF signals is also beneficial because LAADF typically has much higher scattering intensity compared to HAADF, and therefore a greater signal-to-noise ratio can be achieved.

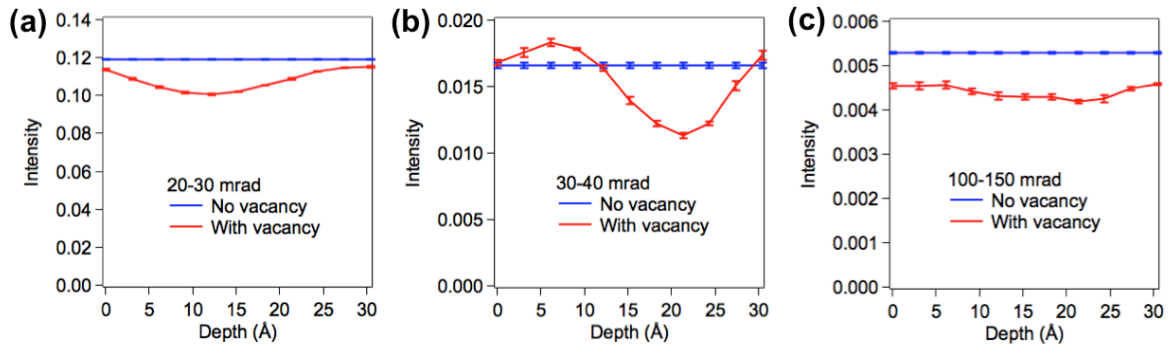

**Figure 5.** Column intensity as a function of vacancy depth position (red curve) compared to the column intensity without a vacancy (blue curve) for detection angles (a) 20-30 mrad, (b) 30-40 mrad, and (c) 100-150 mrad using 20 mrad probe half convergence angle. All simulations used TDS.

3.3. Effect of probe convergence angle

We tested how the high depth sensitivity of 20-40 mrad scattering angles changes when a higher probe half convergence angle, 20 mrad, is used (Fig. 5). The result shows that, while the absolute column intensities change with a higher convergence angle, the overall trend of the intensity versus the depth profile does not significantly vary (Fig. 5a and 5b) in comparison to the profiles with a 9.6 mrad half convergence angle (Fig. 4d and 4e). The trend in HAADF signals with a 100-150 mrad detection range also does not change with respect to the convergence angle, showing limited dependence on the vacancy depth position (Fig. 5c). A similar ripple effect at the 20-40 mrad angle range was also observed in the CBED simulation using a 20 mrad convergence half angle (not shown). The results indicate that the higher probe convergence angle does not significantly affect the trend of the low angle ripple effect and therefore can also be used for vacancy depth imaging.

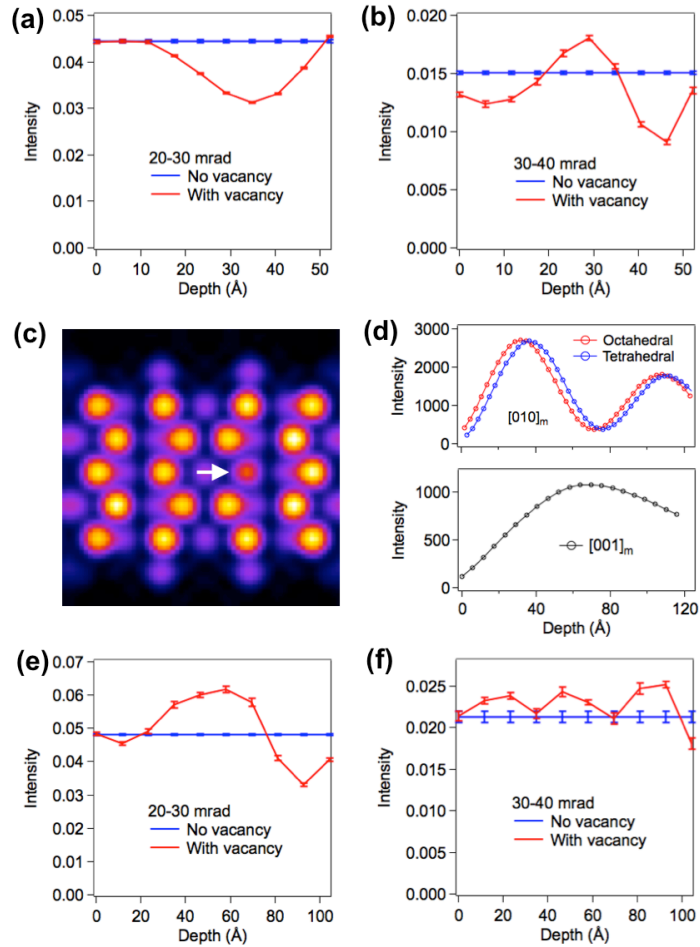

**Figure 6**. Column intensity as a function of vacancy depth position (red curve) compared to the column intensity without a vacancy (blue curve) for detection angles (a) 20-30 mrad and (b) 30-40 mrad, for imaging with $[001]_m$ orientation and sample thickness of 60 Å. (c) shows the image simulated for the vacancy placed at the 9$^{th}$ position (46 Å depth) from the top with a 30-40 mrad detection angle. The arrow indicates the column containing the vacancy. (d) Probe channeling oscillation along (top) $[010]_m$ (bottom) $[001]_m$ orientations in $\beta$-Ga$_2$O$_3$. (e) and (f) are the same simulations as (a) and (b), but with a sample twice as thick (119 Å). For (e) and (f), the intensities for every other vacancy positions were simulated.

3.4. Effect of crystal orientation and thickness

Here we show the vacancy-containing column intensities as compared to the normal column intensities using the other orientation, $[001]_m$. As explained in Section 2, the interatomic distance along the depth direction of the $[001]_m$ orientation (5.8 Å) is nearly twice as long as the $[010]_m$ orientation (3.04 Å). Figure 6 shows the results simulated using the $[001]_m$ orientation with a sample thickness of 60 Å, which is almost twice as thick as the thickness used in the simulation along $[010]_m$ orientation shown in Fig. 4. However, the number of atoms along $[001]_m$ direction in this model is almost equal to the number of atoms along $[010]_m$ direction in the 33 Å thick sample shown in Fig. 4 because of the difference in the interatomic spacing. Despite the significant difference in sample thickness, the intensities simulated using detection ranges 20-30 mrad (Fig. 6a) and 30-40 mrad (Fig. 6b) along the $[001]_m$ orientation show similar trends as compared to the simulations along the $[010]_m$ orientation shown in Fig. 4. This can be explained through the simulation of the probe oscillation shown in Fig. 6d. The oscillation along the $[010]_m$ shows almost linear increase until the first maximum at a depth of ~35 Å, while the oscillation along $[001]_m$ shows the same trend until the maximum at a depth of ~65 Å. This indicates that the number of atoms directly affects the probe channeling oscillation along the depth, not the actual thickness of the sample. Therefore the simulations using 33 Å thick $[010]_m$ and 60 Å thick $[001]_m$ models must have similar de-channeling ripple effects, which results in similar trends in the depth dependence of the vacancy-containing column intensity, despite the differences in thickness. The result also suggests that, depending on which imaging orientation is chosen, the usable sample thickness for 3D vacancy imaging can increase, which is a benefit because thicker TEM samples are easier to prepare in general.

We also performed the same simulation using a 119 Å thick sample along the $[001]_m$ orientation (Figure 6e and 6f). The results show that the depth dependency of the vacancy-containing column intensity becomes rather complicated, especially in 30-40 mrad detection angles. Because of the complicated profile, the 119 Å thickness may be too thick for vacancy depth imaging.

The effect of thicker samples on the contrast of vacancy-containing columns were also tested using the $[010]_m$ orientation. Figure 7 shows the simulation that is identical to the one shown in Fig. 4, but with a thicker sample (55 Å). The intensities of the vacancy-containing columns in both the 20-30 and 30-40 mrad detection ranges show a different and slightly more complex trend as compared to that of the 33 Å thick sample shown in Fig. 4. However, because the trends differ, together the individual profiles of both 10 mrad width detection ranges can be used to determine the vacancy depth position. Therefore, we conclude that the usable sample thickness for the 3D imaging of individual vacancies in $\beta$-$Ga_2O_3$ is about 6 nm, in both the $[010]_m$ and $[001]_m$ orientations.

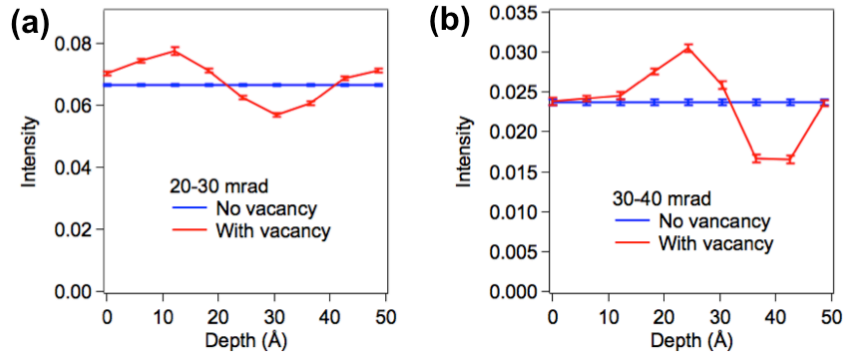

**Figure 7.** Column intensity as a function of vacancy depth position (red curve) compared to the column intensity without a vacancy (blue curve) for detection angles (a) 20-30 mrad, and (b) 30-40 mrad, for imaging with $[010]_m$ orientation and sample thickness of 55 Å. The intensities for every other vacancy position were simulated.

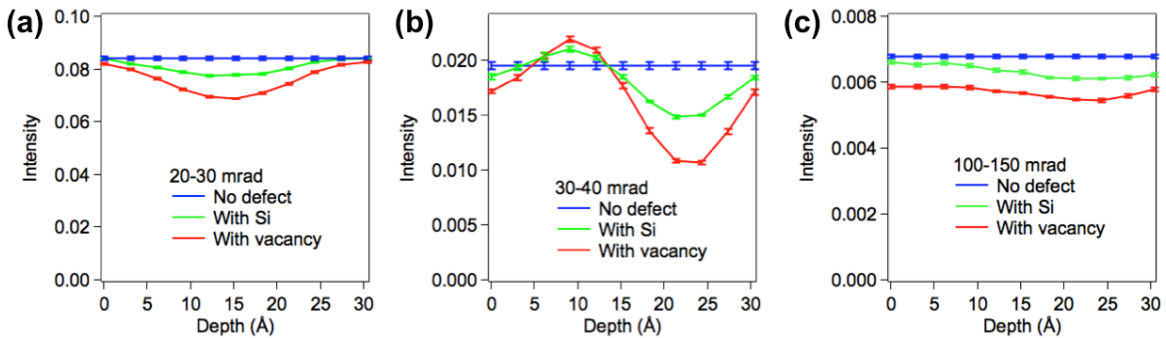

**Figure 8.** Column intensity as a function of Si depth position (red curve) compared to the column intensity without an impurity (blue curve) for detection angles (a) 20-30 mrad, (b) 30-40 mrad, and (c) 100-150 mrad, for imaging with $[010]_m$ orientation. The data with a vacancy (red curve) is the same data presented in Fig. 4. All simulations used TDS and a probe convergence half angle of 9.6 mrad.

3.5. 3D imaging of lighter dopant atoms using LAADF angles

In the previous sections, we have demonstrated that the 3D imaging of vacancies should be possible using selective LAADF angles (20-40 mrad). Here we show that the 3D imaging of lighter dopants with accurate depth information should also be possible using the same selective LAADF angles. We chose Si as the lighter dopant, as it is one of the most commonly used dopants in $\beta$-$Ga_2O_3$ [40, 50], and controlling the unintentional doping of Si in pure $\beta$-$Ga_2O_3$ is also an important issue [38, 51]. Si has a $Z$-number that is less than a half of the host Ga atom ($Z$ = 14 vs. 31). Figure 8 shows that the intensity profile along the $[010]_m$ direction as a function of Si dopant atom depth essentially follows the same trend as the vacancy case that we showed earlier. However, the absolute amplitude of the column intensity change is about half that of the vacancy cases in both the 20-30 and 30-40 mrad ranges. This means that, with the presence of experimental uncertainty, the accurate determination of the Si atom along the depth should be possible, but more challenging. It is also important to point out that, it may not be completely possible to determine whether the impurity is a Si dopant or vacancy, as their intensities in both detection ranges are too close to each other in some cases (for example, the positions at 3-6 Å depth range) as shown in Fig. 8a and 8b. In such cases, incorporating the signal from the 3$^{rd}$ detector can help the distinction between a Si dopant and vacancy. For example, if the 3$^{rd}$ detector is placed in a HAADF range 100-150 mrad, the intensity of the Si-containing column is consistently higher than that of the vacancy-containing column throughout the entire depth range (Fig. 8c).

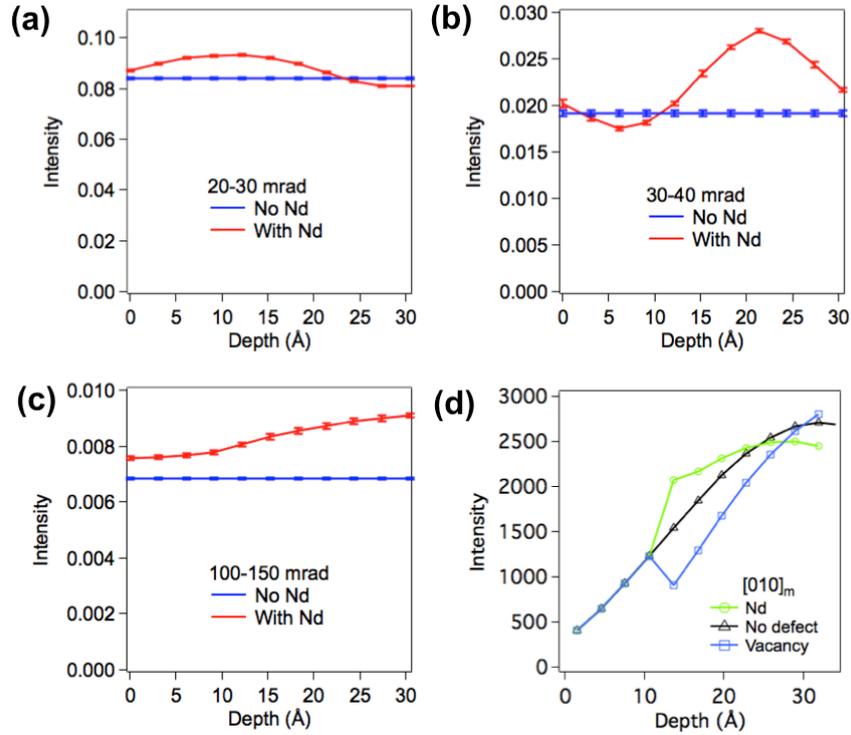

**Figure 9**. Column intensity as a function of Nd dopant depth position (red curve) compared to the column intensity without a vacancy (blue curve) for detection angles (a) 20-30 mrad, and (b) 30-40 mrad, for imaging with $[010]_m$ orientation. (d) Probe channeling intensity profiles along the depth direction without point defects (black) and with point defects, including a vacancy (light blue) and a Nd dopant (light green), placed at the 5$^{th}$ position (13 Å depth) from the top.

3.6. 3D imaging of heavier dopant atoms in $\beta$-Ga$_2$O$_3$

Figure 9 shows the intensity as a function of the depth of a heavier dopant, Nd [43] in $\beta$-Ga$_2$O$_3$. Nd has $Z = 60$, which is about twice as heavy as Ga ($Z = 31$). The overall trends of the intensities in the 20-30 and 30-40 mrad ranges in Fig. 9 are different from the trends shown in Fig. 4 and 8. In comparison to the vacancy and Si cases, the first derivative of the graphs shows the opposite sign. This indicates that the de-channeling ripple effect still happens with heavier dopants but in a different way. The result shows that 3D imaging of heavier dopants should also be possible using the selective LAADF angles. However, the HAADF angle, shown in Fig. 9c, shows monotonic increase as a function of the dopant depth, and therefore the HAADF signal alone should be adequate to determine the dopant depth position. The monotonic increase in the intensity shown in Fig. 9c is consistent to the trend observed in Gd dopant placed in a SrTiO₃ host in the previous work [6]. As we have shown in the previous sections, the monotonic increase of HAADF signal as a function of defect depth does not happen in the vacancy and lighter dopant cases. The origin of this difference can be found in the probe channeling oscillation along the depth shown in Fig. 9d. As pointed out by Hwang et al. [6], the HAADF intensity strongly depends on the probe channeling oscillation along the depth. Here, the probe intensity as a function of depth in a perfect column (black) is compared to those containing Nd (light green) and vacancy (light blue). In the case of a heavier (Nd) point defect, the probe intensity increases at the position of the atom. This means that, while the Nd atom slightly alters the probe channeling profile in the column, the final image intensity will always be an additive sum of the scattering intensities from each atom positions, which result in the monotonic increase of HAADF signals up to the first maximum. However, when a vacancy is placed instead of the heavier dopant, the probe channeling intensity decreases right at the position of the vacancy, and the amount of the decrease is different depending the depth position of the vacancy. This results in a flat or even noisy trend in the final HAADF intensity profile, shown in Fig. 2c and 4f, rather than a monotonic change.

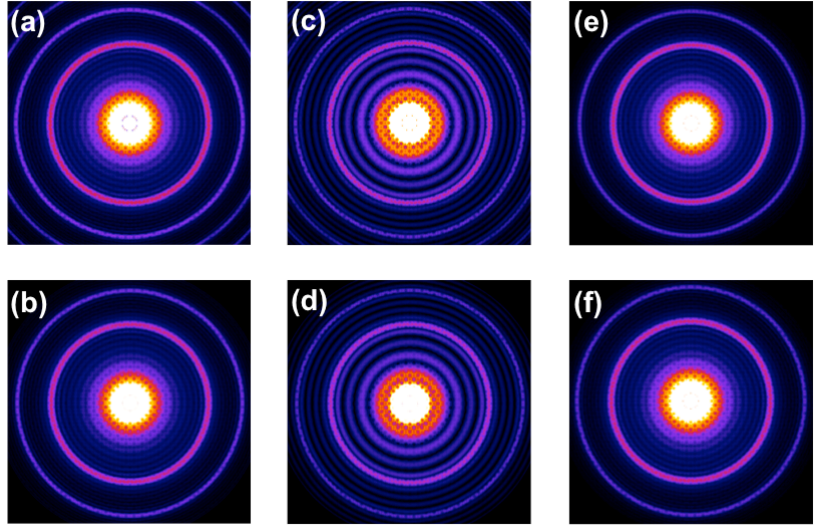

**Figure 10**. CBED simulated for the Sr column in SrTiO₃ with no vacancy, slice thickness = 1.953 Å and (a) 4 × 4 × 8 (in *x*, *y*, and *z*) unit cells and 1024 × 1024 real space pixel sampling, and (b) 20 × 20 × 8 unit cells and 2048 × 2048 pixel sampling. (c) and (d) are the same simulations as (a) and (b), respectively, but with a vacancy positioned in the middle of the column. (e and f) CBED simulated with no vacancy, and slice thickness of (e) 1.3 Å and (f) 0.98 Å. No TDS was used.

3.7. Verification of the low angle ripple effect and vacancy contrast

As previously shown, capturing the signals from the low angle extra ripple effect in the 20-40 mrad angle range can critically assist the 3D imaging of vacancies and lighter atoms. Here we verify the generality of the ripple effect by simulating the CBED patterns using a higher symmetry crystal, $SrTiO_3$, and also showing that the ripple effect is not a caused by any simulation artifacts. Figure 10 shows the CBED patterns simulated without TDS for the Sr position in $SrTiO_3$ (thickness = 19.5 Å) with no vacancy (Fig. 10a) and with a vacancy positioned in the middle of the column (Fig. 10c). The low angle de-channeling ripples become stronger in the CBED with a vacancy, which is consistent to the $β\text{-}Ga_2O_3$ case shown in Fig. 3b. The exact same change in ripples also occurs with a model that is laterally five times larger (Fig. 10b and 10d), indicating that the ripple effect is not an artifact caused by the use of a small real space model. The ripple patterns are also unaffected by the use of a smaller slice thickness in the multislice simulations (Fig. 10e and 10f).

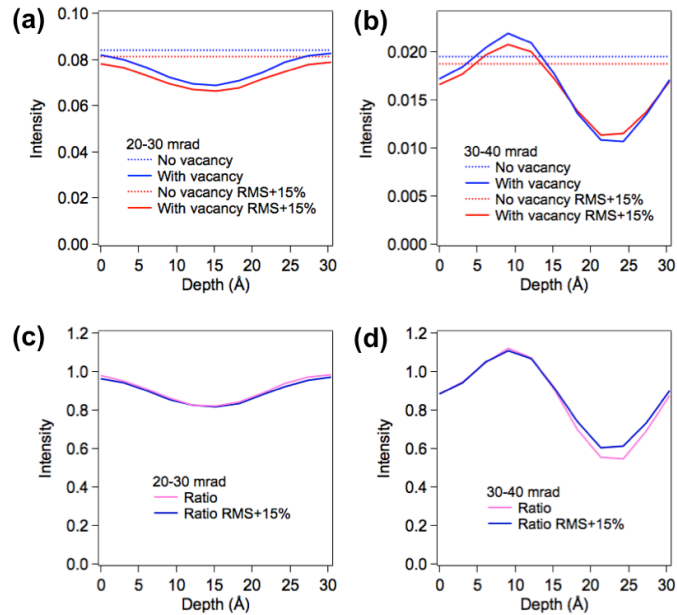

**Figure 11.** Column intensity data with vs. without a vacancy shown in Fig. 4 (solid and dotted blue curves, respectively), as compared to the same simulation run with thermal vibration RMS value increased by 15% (solid and dotted red curves) for (a) 20-30 mrad, and (b) 30-40 mrad. (c) and (d) show the intensity ratio ($\frac{with\ vacancy}{without\ vacancy}$) for the normal RMS (pink) and RMS+15% (blue) data in (a) and (b), respectively.

We also tested the effect of the amplitude of thermal vibration to the contrast of the vacancy-containing column in selective LAADF imaging. As shown in Fig. 3, the ripple effect is present even when TDS is used, which is why the LAADF can distinguish the columns with bulk vacancies. However, we have to question if the RMS value for TDS used in the simulation is really an accurate one. The multislice code assumes isotropic thermal vibration in Cartesian coordinates (or just in $x$ and $y$ directions since the vibration along $z$ will be typically smaller than the slice thickness and therefore becomes irrelevant) [44], but this may be erroneous due to the structural anisotropy of $\beta$-$Ga_2O_3$ [45]. Due to the anisotropic thermal vibration in $\beta$-$Ga_2O_3$ [45], we estimate that the isotropic RMS value that we used may be off by ~15% depending on the imaging orientation. Therefore we tested whether the isotropic assumption could cause systematic error in the column intensity by increasing the thermal vibration RMS values by 15%. The result shows that the absolute column intensity values can decrease by 3-5% (Fig. 11a and 11b) due to the increase in RMS, which indicates that the use of accurate anisotropic TDS parameters will be required for a fully quantitative comparison between experimental and simulated images. Nevertheless, the ratio of the intensities between the columns with and without a vacancy does not change significantly (Fig. 11c and 11d), hence the relative contrast of the vacancy-containing column will not be affected by the potential error.

**4. Discussion**

4.1. Experimental realization of multiple, narrow ADF ranges

We demonstrated that, using small detection angle range in LAADF mode, such as 20-30 and 30-40 mrad ranges, can substantially enhance the contrast of the defect-containing column with its dependence on the defect depth information. Our result can therefore create new opportunities for imaging point defects in material with accurate 3D information, if the experiment uses the same detection angles presented in this work. Experimentally, the best option to achieve such narrow LAADF angle ranges with ~10 mrad width would be using the direct electron detector with ultrafast read-out speed (*e.g.* [52]), which can acquire CBED patterns (such as the one shown in Fig. 3) for each probe position while scanning the probe across the sample area. From the CBED patterns collected for every probe position, reconstructing the STEM images using the signals in multiple scattering angle ranges, for example, 20-30 and 30-40 mrad, and HAADF angles, should be possible. The experiment should also be possible using conventional STEM ADF detectors. The small LAADF angle range can be selected, for example, by using a normal ADF detector combined with the use of the shadow of an objective aperture onto the detector. In our FEI Titan STEM, using a camera length of 230 mm and an objective aperture with 100 μm diameter, the angle range of 20-27 mrad can be selected. In a similar manner, the 30-40 mrad angle range can be selected as well. Simultaneous acquisition of the two images may be difficult, but they can be acquired consecutively, as we have demonstrated in Zhang et al [26].

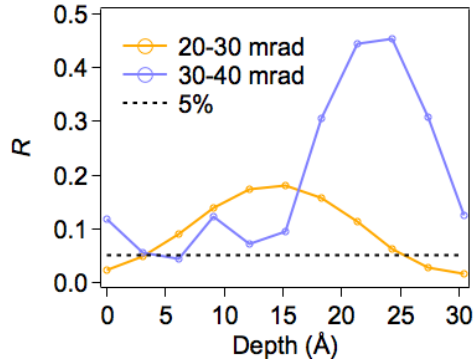

**Figure 12**. The difference ratio $R = |I_{vac} - I_0| / I_0$ between the column intensities with and without a vacancy along [010]$_m$ orientation as a function of vacancy depth. $I_{vac}$ is the intensity of the vacancy-containing column and $I_0$ is the intensity of the column without a vacancy.

4.2. Allowable experimental uncertainties

The determination of the depth of point defects will be ultimately limited by the experimental uncertainties in real imaging experiments [6]. Here we have estimated the maximum allowed experimental uncertainty based on the vacancy depth imaging case shown in Fig. 4. For 20-30 and 30-40 mrad data, we have calculated the difference ratio ($R$) between the column intensities with and without a vacancy using $R = |I_{vac} - I_0| / I_0$, where $I_{vac}$ is the intensity of the vacancy-containing column and $I_0$ is the intensity of the column without vacancy. The $R$ value changes as a function of the vacancy depth as displayed in Fig. 12. For example, for the vacancy placed at the 5$^{th}$ position from the top, the $R$ value is 0.174 for the 20-30 mrad range, and 0.071 for the 30-40 mrad range. This means that if the experimental uncertainty is less than 0.071 (~7%), the accurate determination of the vacancy located at the 5$^{th}$ position should be possible. Throughout the whole thickness range, at least one of the 20-30 and 30-40 mrad data points is above the 5% uncertainty line (dashed line), indicating that the exact vacancy depth identification should be possible if the experimental uncertainty (in standard deviation) can be kept less than 5%. 5% is about half the uncertainty we previously reported with HAADF dopant imaging [6]. Since the LAADF detection angles have much higher signal, they should increase the signal-to-noise ratio, and if we assume that the experimental uncertainty only depends on the signal-to-noise ratio in the signal collection, the experimental error should decrease. However, the non-ideal sample conditions, such as surface roughness, can also contribute to the experimental uncertainty. This may be especially important because LAADF intensity is known to be more sensitive to surface features, although it is unclear whether this will still be the case when the narrower LAADF angle ranges are selected. Surface reconstruction may also complicate the image contrast, but no surface reconstruction in $\beta$-Ga$_2$O$_3$ has been reported [51, 53].

We must also note that the present work does not consider the relaxation or distortion of the atom positions that can potentially occur near the point defects. Such atomic displacements can also slightly alter the electron channeling, but it is difficult to guess how exactly they would occur prior to the experimental imaging. Therefore, while our results provide the initial guidelines to the optical and sample parameters for the point defect imaging, an additional simulation including the atomic displacement information may be required after experimental imaging to fully understand the details of the point defect structure.

## 5. Conclusion

In summary, we have shown that the 3D characterization of vacancies, lighter and heavier dopants can be determined from quantitative STEM imaging by optimizing the ADF detection angles.

The selection of a small range of LAADF signals can make the contrast of vacancy and lighter dopant-containing atomic columns more depth dependent, while HAADF signals are more sensitive to the depth of heavier dopants. Simulated CBED patterns revealed the cause of the LAADF's dependence on the vacancy depth to be a ripple effect, which is the result of de-channeling of the electron due to the existence of the vacancy in the column. Capturing the de-channeling signal with a narrowly selected range of ADF angles critically assists the 3D imaging of vacancies and lighter atoms by enhancing the contrast of the point defect-containing columns. Using this method, we have shown that the probe convergence angle does not significantly change the overall trend of the profile and that depending on crystal orientation, thicker TEM foils can be used. However, the application of this method to determine the depth of point defects is ultimately limited by the experimental uncertainties in real imaging experiments. With the use of the detection angles presented, new opportunities for imaging point defects in materials with accurate 3D information can be created.

**Acknowledgement**

The authors acknowledge the use of the computing facilities at the Ohio Supercomputer Center, which were used to carry out the STEM image and diffraction simulations. S.I. was supported by the Seed Grant from the Institute for Materials Research at the Ohio State University and the Center for Emergent Materials, an NSF funded MRSEC under award DMR-1420451

# References

1. G.P. Lansbergen, R. Rahman, C.J. Wellard, I. Woo, J. Caro, N. Collaert, S. Biesemans, G. Klimeck, L.C.L. Hollenberg, and S. Rogge, Gate-induced quantum-confinement transition of a single dopant atom in a silicon FinFET. *Nat Phys*, **4**(8), 656-661 (2008).
2. P.M. Koenraad and M.E. Flatte, Single dopants in semiconductors. *Nat Mater*, **10**(2), 91-100 (2011).
3. R. Krause-Rehberg and S.L. Hartmut, Positron Annihilation in Semiconductors. (1999).
4. R.F. Loane, E.J. Kirkland, and J. Silcox, Visibility of single heavy atoms on thin crystalline silicon in simulated annular dark-field STEM images. *Acta Crystallographica Section A*, **44**(6), 912-927 (1988).
5. P.M. Voyles, D.A. Muller, J.L. Grazul, P.H. Citrin, and H.J.L. Gossmann, Atomic-scale imaging of individual dopant atoms and clusters in highly n-type bulk Si. *Nature*, **416**(6883), 826-829 (2002).
6. J. Hwang, J.Y. Zhang, A.J. D'Alfonso, L.J. Allen, and S. Stemmer, Three-Dimensional Imaging of Individual Dopant Atoms in $SrTiO_3$. *Physical Review Letters*, **111**(26), 266101 (2013).
7. S.H. Oh, K.v. Benthem, S.I. Molina, A.Y. Borisevich, W. Luo, P. Werner, N.D. Zakharov, D. Kumar, S.T. Pantelides, and S.J. Pennycook, Point Defect Configurations of Supersaturated Au Atoms Inside Si Nanowires. *Nano Letters*, **8**(4), 1016-1019 (2008).
8. H. Okuno, J.-L. Rouvière, P.-H. Jouneau, P. Bayle-Guillemaud, and B. Daudin, Visualization of Tm dopant atoms diffused out of GaN quantum dots. *Applied Physics Letters*, **96**(25), 251908 (2010).
9. M. Bar-Sadan, J. Barthel, H. Shtrikman, and L. Houben, Direct Imaging of Single Au Atoms Within GaAs Nanowires. *Nano Letters*, **12**(5), 2352-2356 (2012).
10. A. Mittal and K.A. Mkhoyan, Limits in detecting an individual dopant atom embedded in a crystal. *Ultramicroscopy*, **111**(8), 1101-1110 (2011).
11. M. Varela, S.D. Findlay, A.R. Lupini, H.M. Christen, A.Y. Borisevich, N. Dellby, O.L. Krivanek, P.D. Nellist, M.P. Oxley, L.J. Allen, and S.J. Pennycook, Spectroscopic Imaging of Single Atoms Within a Bulk Solid. *Physical Review Letters*, **92**(9), 095502 (2004).
12. A.A. Gunawan, K.A. Mkhoyan, A.W. Wills, M.G. Thomas, and D.J. Norris, Imaging "Invisible" Dopant Atoms in Semiconductor Nanocrystals. *Nano Letters*, **11**(12), 5553-5557 (2011).
13. M.D. Rossell, Q.M. Ramasse, S.D. Findlay, F. Rechberger, R. Erni, and M. Niederberger, Direct Imaging of Dopant Clustering in Metal–Oxide Nanoparticles. *ACS Nano*, **6**(8), 7077-7083 (2012).
14. G.-z. Zhu, S. Lazar, A.P. Knights, and G.A. Botton, Atomic-level 2-dimensional chemical mapping and imaging of individual dopants in a phosphor crystal. *Physical Chemistry Chemical Physics*, **15**(27), 11420-11426 (2013).
15. H.L. Xin and D.A. Muller, Three-Dimensional Imaging in Aberration-Corrected Electron Microscopes. *Microscopy and Microanalysis*, **16**(04), 445-455 (2010).

16. R. Ishikawa, A.R. Lupini, S.D. Findlay, T. Taniguchi, and S.J. Pennycook, Three-Dimensional Location of a Single Dopant with Atomic Precision by Aberration-Corrected Scanning Transmission Electron Microscopy. *Nano Letters*, **14**(4), 1903-1908 (2014).
17. R. Ishikawa, A.R. Lupini, Y. Hinuma, and S.J. Pennycook, Large-angle illumination STEM: Toward three-dimensional atom-by-atom imaging. *Ultramicroscopy*, **151**, 122-129 (2015).
18. K. van Benthem, A.R. Lupini, M.P. Oxley, S.D. Findlay, L.J. Allen, and S.J. Pennycook, Three-dimensional ADF imaging of individual atoms by through-focal series scanning transmission electron microscopy. *Ultramicroscopy*, **106**(11–12), 1062-1068 (2006).
19. P. Wang, A.J. D'Alfonso, A. Hashimoto, A.J. Morgan, M. Takeguchi, K. Mitsuishi, M. Shimojo, A.I. Kirkland, L.J. Allen, and P.D. Nellist, Contrast in atomically resolved EF-SCEM imaging. *Ultramicroscopy*, **134**, 185-192 (2013).
20. P. Wang, A. Hashimoto, M. Takeguchi, K. Mitsuishi, M. Shimojo, Y. Zhu, M. Okuda, A.I. Kirkland, and P.D. Nellist, Three-dimensional elemental mapping of hollow $Fe_2O_3@SiO_2$ mesoporous spheres using scanning confocal electron microscopy. *Applied Physics Letters*, **100**(21), 213117 (2012).
21. H.L. Xin, C. Dwyer, D.A. Muller, H. Zheng, and P. Ercius, Scanning Confocal Electron Energy-Loss Microscopy Using Valence-Loss Signals. *Microscopy and Microanalysis*, **19**(04), 1036-1049 (2013).
22. A.R. Lupini, A.Y. Borisevich, J.C. Idrobo, H.M. Christen, M. Biegalski, and S.J. Pennycook, Characterizing the Two- and Three-Dimensional Resolution of an Improved Aberration-Corrected STEM. *Microscopy and Microanalysis*, **15**(05), 441-453 (2009).
23. M.M.J. Treacy and J.M. Gibson, Coherence and multiple scattering in "Z-contrast" images. *Ultramicroscopy*, **52**(1), 31-53 (1993).
24. P.M. Voyles, D.A. Muller, and E.J. Kirkland, Depth-Dependent Imaging of Individual Dopant Atoms in Silicon. *Microscopy and Microanalysis*, **10**(02), 291-300 (2004).
25. L.F. Kourkoutis, M.K. Parker, V. Vaithyanathan, D.G. Schlom, and D.A. Muller, Direct measurement of electron channeling in a crystal using scanning transmission electron microscopy. *Physical Review B*, **84**(7), 075485 (2011).
26. J.Y. Zhang, J. Hwang, B.J. Isaac, and S. Stemmer, Variable-angle high-angle annular dark-field imaging: application to three-dimensional dopant atom profiling. *Scientific Reports*, **5**, 12419 (2015).
27. P.M. Voyles, J.L. Grazul, and D.A. Muller, Imaging individual atoms inside crystals with ADF-STEM. *Ultramicroscopy*, **96**(3–4), 251-273 (2003).
28. J.M. LeBeau, S.D. Findlay, L.J. Allen, and S. Stemmer, Quantitative Atomic Resolution Scanning Transmission Electron Microscopy. *Physical Review Letters*, **100**(20), 206101 (2008).
29. J.M. LeBeau, S.D. Findlay, L.J. Allen, and S. Stemmer, Position averaged convergent beam electron diffraction: Theory and applications. *Ultramicroscopy*, **110**(2), 118-125 (2010).
30. H.H. Tippins, Optical Absorption and Photoconductivity in the Band Edge of β-$Ga_2O_3$. *Physical Review*, **140**(1A), A316-A319 (1965).

31. J.B. Varley, J.R. Weber, A. Janotti, and C.G. Van de Walle, Oxygen vacancies and donor impurities in β-Ga2O3. *Applied Physics Letters*, **97**(14), 142106 (2010).
32. N. Ueda, H. Hosono, R. Waseda, and H. Kawazoe, Synthesis and control of conductivity of ultraviolet transmitting β-Ga2O3 single crystals. *Applied Physics Letters*, **70**(26), 3561-3563 (1997).
33. M. Higashiwaki, K. Sasaki, A. Kuramata, T. Masui, and S. Yamakoshi, Gallium oxide (Ga2O3) metal-semiconductor field-effect transistors on single-crystal β-Ga2O3 (010) substrates. *Applied Physics Letters*, **100**(1), 013504 (2012).
34. T. Oishi, Y. Koga, K. Harada, and M. Kasu, High-mobility β-Ga2O3($\bar{2}01$) single crystals grown by edge-defined film-fed growth method and their Schottky barrier diodes with Ni contact. *Applied Physics Express*, **8**(3), 031101 (2015).
35. M. Orita, H. Ohta, M. Hirano, and H. Hosono, Deep-ultraviolet transparent conductive β-Ga2O3 thin films. *Applied Physics Letters*, **77**(25), 4166-4168 (2000).
36. M. Higashiwaki, K. Sasaki, T. Kamimura, M. Hoi Wong, D. Krishnamurthy, A. Kuramata, T. Masui, and S. Yamakoshi, Depletion-mode Ga2O3 metal-oxide-semiconductor field-effect transistors on β-Ga2O3 (010) substrates and temperature dependence of their device characteristics. *Applied Physics Letters*, **103**(12), 123511 (2013).
37. Z. Zhang, E. Farzana, A.R. Arehart, and S.A. Ringel, Deep level defects throughout the bandgap of (010) β-Ga2O3 detected by optically and thermally stimulated defect spectroscopy. *Applied Physics Letters*, **108**(5), 052105 (2016).
38. J.B. Varley, H. Peelaers, A. Janotti, and C.G.V.d. Walle, Hydrogenated cation vacancies in semiconducting oxides. *Journal of Physics: Condensed Matter*, **23**(33), 334212 (2011).
39. T. Zacherle, P.C. Schmidt, and M. Martin, Ab initio calculations on the defect structure of Ga2O3. *Physical Review B*, **87**(23), 235206 (2013).
40. E. Korhonen, F. Tuomisto, D. Gogova, G. Wagner, M. Baldini, Z. Galazka, R. Schewski, and M. Albrecht, Electrical compensation by Ga vacancies in Ga2O3 thin films. *Applied Physics Letters*, **106**(24), 242103 (2015).
41. H. Li and J. Robertson, Behaviour of hydrogen in wide band gap oxides. *Journal of Applied Physics*, **115**(20), 203708 (2014).
42. J.B. Varley, A. Janotti, C. Franchini, and C.G. Van de Walle, Role of self-trapping in luminescence and $p$-type conductivity of wide-band-gap oxides. *Physical Review B*, **85**(8), 081109 (2012).
43. Z. Wu, G. Bai, Q. Hu, D. Guo, C. Sun, L. Ji, M. Lei, L. Li, P. Li, J. Hao, and W. Tang, Effects of dopant concentration on structural and near-infrared luminescence of Nd3+-doped beta-Ga2O3 thin films. *Applied Physics Letters*, **106**(17), 171910 (2015).
44. E.J. Kirkland, R.F. Loane, and J. Silcox, Simulation of annular dark field stem images using a modified multislice method. *Ultramicroscopy*, **23**(1), 77-96 (1987).
45. J. Ahman, G. Svensson, and J. Albertsson, A Reinvestigation of [beta]-Gallium Oxide. *Acta Crystallographica Section C*, **52**(6), 1336-1338 (1996).
46. J. Taylor, An Introduction to Error Analysis, The Study of Uncertainties in Physical Measurements.).
47. H. E, K.E. MacArthur, T.J. Pennycook, E. Okunishi, A.J. D'Alfonso, N.R. Lugg, L.J. Allen, and P.D. Nellist, Probe integrated scattering cross sections in the analysis of atomic resolution HAADF STEM images. *Ultramicroscopy*, **133**, 109-119 (2013).

48. D.A. Muller, N. Nakagawa, A. Ohtomo, J.L. Grazul, and H.Y. Hwang, Atomic-scale imaging of nanoengineered oxygen vacancy profiles in SrTiO3. *Nature*, **430**(7000), 657-661 (2004).
49. L. Fitting, S. Thiel, A. Schmehl, J. Mannhart, and D.A. Muller, Subtleties in ADF imaging and spatially resolved EELS: A case study of low-angle twist boundaries in SrTiO3. *Ultramicroscopy*, **106**(11–12), 1053-1061 (2006).
50. E.G. Víllora, K. Shimamura, Y. Yoshikawa, T. Ujiie, and K. Aoki, Electrical conductivity and carrier concentration control in β-Ga2O3 by Si doping. *Applied Physics Letters*, **92**(20), 202120 (2008).
51. K. Iwaya, R. Shimizu, H. Aida, T. Hashizume, and T. Hitosugi, Atomically resolved silicon donor states of β-Ga2O3. *Applied Physics Letters*, **98**(14), 142116 (2011).
52. G. McMullan, A.R. Faruqi, D. Clare, and R. Henderson, Comparison of optimal performance at 300 keV of three direct electron detectors for use in low dose electron microscopy. *Ultramicroscopy*, **147**, 156-163 (2014).
53. T.C. Lovejoy, E.N. Yitamben, N. Shamir, J. Morales, E.G. Villora, K. Shimamura, S. Zheng, F.S. Ohuchi, and M.A. Olmstead, Surface morphology and electronic structure of bulk single crystal β-Ga2O3(100). *Applied Physics Letters*, **94**(8), 081906 (2009).
28